\documentclass[sigconf]{acmart}

\usepackage[linesnumbered,ruled,vlined]{algorithm2e}

\copyrightyear{2026}
\acmYear{2026}
\setcopyright{cc}
\setcctype{by}
\acmConference[RecSys '26]{20th ACM Conference on Recommender Systems}{September 27-October 02, 2026}{Minneapolis, MN, USA}
\acmBooktitle{20th ACM Conference on Recommender Systems (RecSys '26), September 27-October 02, 2026, Minneapolis, MN, USA}
\acmDOI{10.1145/3773078.3831870}
\acmISBN{979-8-4007-2284-4/2026/09}

\begin{document}

\title{Progressive Alignment of Recommender Foundation Model through Multi-Phase Post-Training}

\author{Oseong Choi}
\authornote{Equal contribution.}
\email{oseong.choi@webtoonscorp.com}
\orcid{0009-0009-3462-6689}
\affiliation{%
  \institution{NAVER WEBTOON}
  \city{Seongnam-si}
  \country{Republic of Korea}
}

\author{Hoeinn Kim}
\authornotemark[1]
\email{hine8648@webtoonscorp.com}
\orcid{0009-0006-6865-0511}
\affiliation{%
  \institution{NAVER WEBTOON}
  \city{Seongnam-si}
  \country{Republic of Korea}
}

\author{Jihoon Lee}
\email{zhoon.lee@webtoonscorp.com}
\orcid{0009-0002-0334-6771}
\affiliation{%
  \institution{NAVER WEBTOON}
  \city{Seongnam-si}
  \country{Republic of Korea}
}

\author{Byungsoo Kang}
\email{bsoo414@webtoonscorp.com}
\orcid{0000-0002-7359-0713}
\affiliation{%
  \institution{NAVER WEBTOON}
  \city{Seongnam-si}
  \country{Republic of Korea}
}

\author{Taeyeong Jang}
\email{teo.jang@webtoonscorp.com}
\orcid{0009-0005-6692-0525}
\affiliation{%
  \institution{NAVER WEBTOON}
  \city{Seongnam-si}
  \country{Republic of Korea}
}

\renewcommand{\shortauthors}{Choi and Kim, et al.}

\begin{abstract}
Foundation model(FM) for recommendation has shown strong ability to model long-horizon sequential user behavior. In practice, a single pretrained foundation model is often adapted to diverse downstream serving surfaces through Supervised Fine-Tuning(SFT). However, optimizing task-specific objectives such as clicks or likes does not necessarily align the serving policy with the business metrics that determine recommendation quality.

We propose a three-phase progressive post-training framework that explicitly separates downstream adaptation from business-metric alignment. The adaptation stage is decomposed into Linear Probing(LP) and Full Fine-Tuning(FFT): LP first stabilizes randomly initialized downstream heads within a frozen pretrained representation space, and FFT then jointly specializes the full model for the target task. On top of this stabilized policy, Reinforcement Fine-Tuning(RFT) aligns the model with practical business objectives using a learned reward model. Rather than directly optimizing the serving policy on sparse business targets, we train the policy on dense implicit feedback and use business-metric supervision only for reward modeling.

Offline experiments show that the progressive LP-FFT-RFT framework outperforms single-phase alternatives, and that reward-based alignment yields a stronger serving policy than directly using the reward model itself for ranking. Large-scale online A/B tests further show that the proposed framework improves production recommendation quality over a conventional non-foundation baseline. A reference implementation is available at \url{https://github.com/webtoon/rec-fm-progressive-alignment}.
\end{abstract}




\maketitle

\section{Introduction}
Foundation models have significantly advanced sequential recommendation by utilizing large-scale autoregressive pretraining to capture complex user behavioral patterns. In practice, a single pretrained FM is fine-tuned via SFT and deployed across a wide variety of serving surfaces, each associated with distinct task objectives. However, the task-level objectives optimized by SFT (e.g., predicting clicks or likes) do not always coincide with the business metrics that determine actual recommendation quality. Even when the training objective is designed to directly target such metrics, the underlying supervision signals like purchase are inherently sparse, making it difficult for the fine-tuned model to achieve sufficient generalization. Closing this gap requires explicitly aligning the policy with business metrics through a reward-driven post-training process.

A representative approach to such reward-driven post-training is RFT, which optimizes the policy against a reward signal derived from human or implicit feedback. In Large Language Models(LLMs), RFT has achieved notable success by directly incorporating human feedback to align pretrained models with desired behaviors. Adapting this approach to recommender systems, however, presents two distinct challenges.

First, directly fine-tuning the massive FM backbone alongside randomly initialized task-specific modules in a single step induces catastrophic forgetting of the pretrained representations. High-variance gradients originating from the uninitialized modules can destabilize the backbone before the downstream components have learned meaningful representations. A structured adaptation process is therefore required to integrate task-specific knowledge while preserving the FM’s pretrained behavioral representations.

Second, aligning the serving policy with actual business metrics is non-trivial. A straightforward approach is to directly optimize for business metrics such as long-term retention or purchase conversion. However, supervision for these objectives is inherently sparse and delayed, limiting the generalization of the resulting policy across the full candidate space. In practice, SFT is therefore applied to dense implicit feedback (e.g., clicks, dwell time) to build a base policy with strong ranking capacity, but this leaves a gap between the task-level objective and the target business metric.

Although such business metrics can also be modeled directly, the resulting model is not necessarily the best deployable serving policy. Because long-term business labels are much sparser than dense implicit feedback, the learned scores often capture the direction of utility without providing sufficient discrimination for large-scale ranking. By contrast, dense implicit signals provide richer supervision over candidate items and therefore produce policies with stronger ranking capacity. This motivates a separation of roles: we use dense implicit feedback to train the serving policy and use business-metric supervision to train a reward model that provides an alignment signal.

To address these challenges, we propose a three-phase progressive post-training framework that explicitly separates task-specific knowledge adaptation from business metric alignment. The first two phases constitute the knowledge adaptation stage. LP first aligns the semantic spaces of the pretrained FM and the downstream task modules by optimizing only the downstream components within the frozen FM representation space. FFT then unfreezes all parameters, allowing the model to further specialize for the target task while preserving the representations consolidated during LP. The third phase performs business metric alignment via RFT. We formulate the target business metric as a reward model learned from logged interaction data and use it as an alignment signal rather than as the primary serving objective. Rather than deploying the reward model directly as the serving policy, we use it to fine-tune a policy trained on dense implicit feedback, thereby steering its ranking behavior toward long-term business objectives while preserving the discrimination established during SFT.

Offline experiments demonstrate that our three-phase progressive framework achieves superior performance compared to single-phase SFT, with each phase contributing incrementally to the final performance. Furthermore, aligning a policy trained on dense implicit signals via RFT outperforms both directly supervising a serving policy on sparse business-metric labels and deploying the reward model itself as the serving policy. Online A/B testing in a large-scale production environment confirms that our framework successfully optimizes for long-term user satisfaction.

\section{Related Work}

\subsection{Foundation Models for Recommender Systems}
Recent advances in sequential recommendation have evolved from task-specific architectures toward large-scale foundation models. Early autoregressive sequential recommenders \cite{hidasi2016gru4rec, kang2018sasrec, rajput2023generative} established next-item prediction as a dominant paradigm for modeling user behavior dynamics. Building upon this autoregressive formulation, subsequent studies revealed that sequential recommendation models exhibit scaling-law behavior, where increasing model capacity and training data consistently improves representation quality and long-horizon behavioral understanding \cite{zhang2024scaling}. These findings marked a critical transition: large sequential recommenders began to function not merely as task-specific rankers but as general-purpose behavioral foundation models pretrained on massive user interaction logs~\cite{jiaqu2024hstu, xiangyi2025pinfm}.

As pretrained sequential recommenders matured, research focus shifted from architectural approach toward downstream adaptation. Rather than training independent models for each recommendation surface, recent work explores transferring pretrained sequential representations across diverse tasks such as ranking, retrieval and cross-domain personalization \cite{gong2023unifiedsr, liang2025externalfm}. Industrial systems further operationalize this paradigm through foundation–expert frameworks  \cite{dai2025realizing}, where a shared foundation model provides universal user representations while lightweight task-specific modules specialize for individual objectives. Large-scale deployments such as LFM4Ads \cite{shangyu2025lfm4ads} extend this idea by enabling multi-granularity transfer of user, item, and user–item interaction representations, supporting feature, module and model-level adaptation across heterogeneous recommendation scenarios. These advances position foundation models not merely as pretrained encoders but as reusable behavioral priors enabling scalable downstream specialization.

\subsection{Post-Training and Alignment of Foundation Models}
Adapting pretrained foundation models to downstream tasks typically involves either training only task-specific heads—often referred to as LP—or conducting FFT across all model parameters. Recent empirical studies have demonstrated that a sequential optimization strategy—specifically, performing LP followed by FFT \cite{kumar2022finetuning}—yields superior performance. To manage stability and computational overhead, parameter-efficient fine-tuning methods such as LoRA \cite{hu2021lora} have been widely utilized. By initializing the downstream heads first, the LP-FT paradigm safely introduces new task-specific datasets without heavily distorting the foundational backbone from the outset.

However, even with these structural strategies, standard SFT remains highly susceptible to catastrophic forgetting. When foundation models are heavily updated to fit specific short-term objectives, they frequently degrade their previously acquired, generalized pretrained knowledge \cite{ramasesh2020anatomy, yunluo2025empiricalcf}. To address this, recent research highlights the advantages of RFT. Recent empirical studies reveal that RFT inherently mitigates catastrophic forgetting; the gradient updates in RFT act as a data-dependent regularizer, effectively preserving prior knowledge and maintaining representation stability far better than standard SFT \cite{lai2025rftgood}.

Originating from LLMs, methods such as Proximal Policy Optimization(PPO) \cite{schulman2017ppo} rely on a parameterized reward model and a separate value network to provide dense scalar feedback. To alleviate the substantial computational overhead of PPO, recent advancements like Group Relative Policy Optimization(GRPO) \cite{shao2024deepseekmath} eliminate the need for a value model by estimating the baseline through relative scoring within a group of sampled outputs. Alternatively, Direct Preference Optimization(DPO) \cite{rafailov2023dpo} bypasses explicit reward modeling entirely by optimizing the policy directly over pairwise preferences. These approaches are commonly instantiated within the Reinforcement Learning from Human Feedback(RLHF) framework \cite{ouyang2022rlhf}. In this paradigm, a reward or preference objective is constructed from explicitly collected human annotations, enabling policy optimization toward desired behaviors while strictly constraining representation drift via KL-divergence penalties.

\section{Methodology}
\label{sec:methodology}

\begin{figure}[t]
    \centering
    \includegraphics[width=0.9\columnwidth]{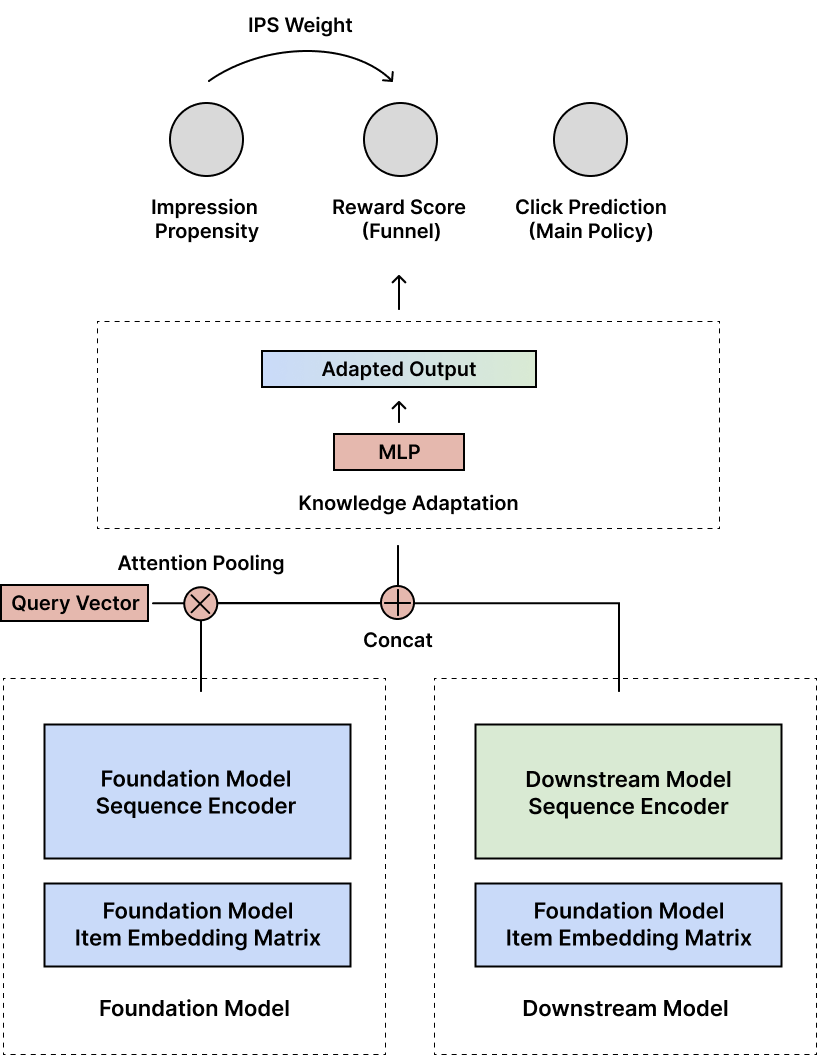}
    \caption{Overall architecture of the foundation–expert framework. The foundation model representation is fused with the downstream task-specific representation to predict engagement, ordinal reward, and impression propensity.}
    \Description{Block diagram of the foundation-expert architecture. A pretrained foundation model sequence encoder and a lightweight downstream sequence encoder each produce a user representation. The two representations are concatenated into a fused representation, which feeds three prediction heads: a serving policy head for engagement, an ordinal reward head for funnel depth, and an impression propensity head.}
    \label{fig:model_arch}
\end{figure}

While a pretrained foundation model provides general-purpose behavioral representations, downstream post-training still involves two practical issues. First, task-specific components and signals should be incorporated without destabilizing the pretrained backbone. Second, dense implicit-feedback supervision can learn a strong serving policy but does not directly optimize the sparse and delayed business metrics that determine practical recommendation quality. We therefore use a three-phase post-training framework: LP, FFT and RFT. The first two phases stabilize downstream adaptation, and the last phase aligns the policy with business objectives.

We first describe the model architecture, where the pretrained FM representation is combined with a lightweight task-specific module for downstream adaptation and a reward model maps sparse business outcomes into dense scalar from logged interaction data. We then present the multi-phase optimization procedure: LP and FFT adapt the fused model to the downstream task while preserving pretrained knowledge and RFT aligns the stabilized serving policy with long-term business objectives using the learned reward signal.

Throughout this paper, we use \emph{FM backbone} to denote the pretrained sequence encoder and \emph{serving policy} to denote the policy head that produces final ranking scores from the fused user representation.

\begin{figure*}[t]
    \centering
    \includegraphics[width=0.95\textwidth]{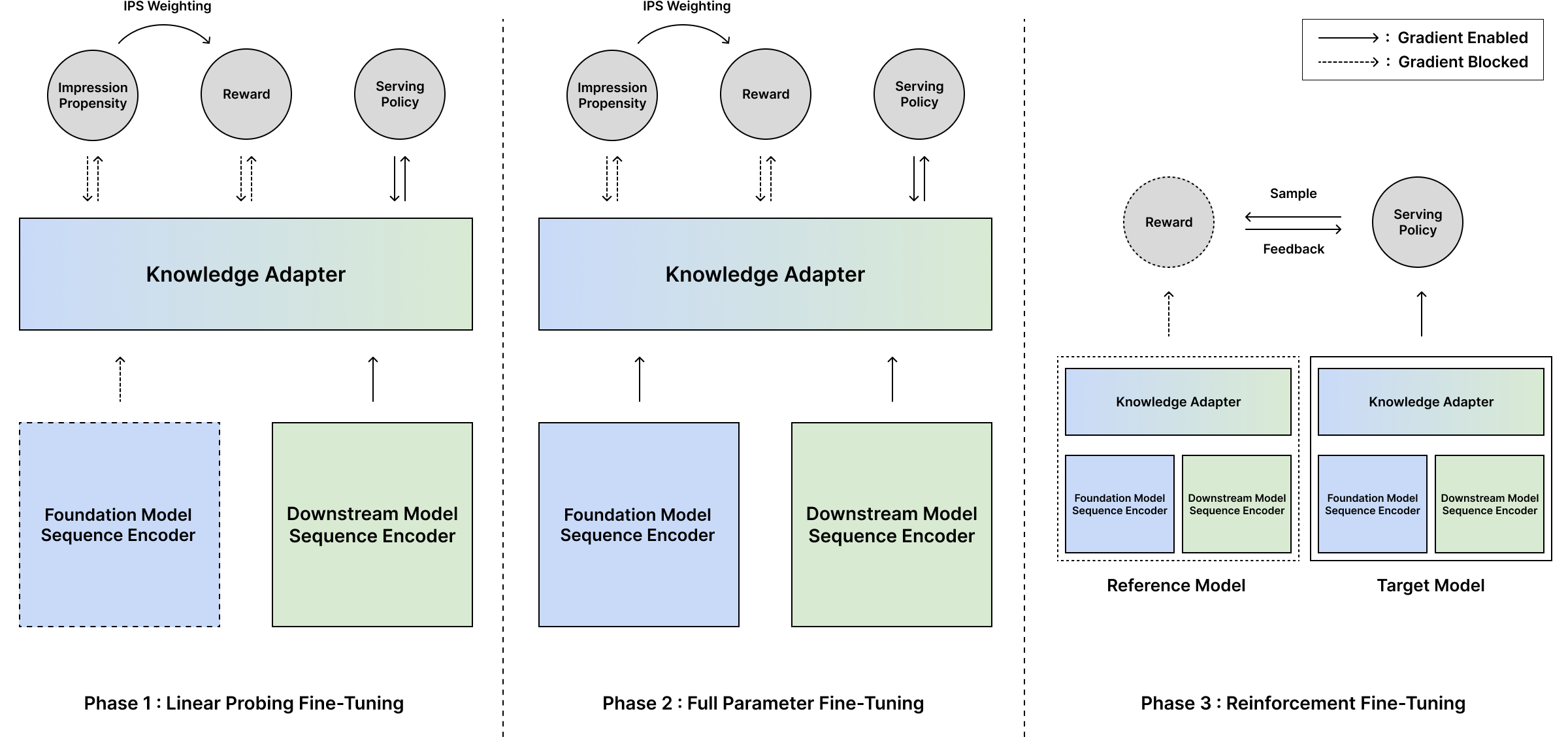}
    \caption{Three-phase progressive post-training pipeline for foundation recommenders. Solid lines indicate gradient flow and dashed lines denote frozen or detached parameters.}
    \Description{Pipeline diagram with three sequential phases. In phase one, linear probing, the foundation model is frozen and only the downstream module and prediction heads receive gradients. In phase two, full fine-tuning, all parameters are unfrozen and updated jointly. In phase three, reinforcement fine-tuning, the serving policy and FM backbone are updated against a frozen reward model and a frozen reference policy, with dashed lines marking the frozen components.}
    \label{fig:model_pipeline}
\end{figure*}

\subsection{Model Architecture}

\subsubsection{Downstream Encoder}

To adapt the pretrained FM to individual serving surfaces under a foundation--expert paradigm, we introduce a lightweight task-specific module that encodes additional surface-specific features and fuses them with the FM representation. Given a user interaction sequence $s$, the pretrained foundation sequence encoder produces a sequence of embeddings,
\[
E = \text{SequenceEncoder}(s)
\]
We then apply cross-attention pooling with a learnable query vector $q$ to compress this variable-length sequence into a fixed-dimensional representation,
\[
\alpha = \mathrm{softmax}\!\left(\frac{q E^\top}{\sqrt{d}}\right), \qquad
h_{\text{fm}} = \alpha E
\]

In parallel, the downstream module produces a representation $h_{\text{ds}}$ from the additional surface-specific features. The two representations are concatenated to form the final user representation,
\[
h = [h_{\text{fm}}; h_{\text{ds}}]
\]
which is then passed to the task heads. This design provides a simple interface for incorporating downstream-specific signals when needed, while keeping the FM representation as the primary backbone of the model.

\subsubsection{Reward Model}

To capture business objectives from logged interaction data without explicit human feedback, we formulate a reward model that predicts the depth of user engagement with a content item. In our Webtoon platform, user engagement can be represented as a multi-stage funnel that reflects progressively deeper content consumption. Based on this structure, we define the reward target as the deepest funnel stage reached by a user for a given item.

In our production setting, we instantiate this target as a six-level funnel ranging from impression to paid-content completion. We model the funnel using ordinal regression~\cite{buyl2023rankformer}. Let $f$ denote the observed funnel depth for a user--item pair $(u,v)$. We define $C=5$ cutpoints and construct ordinal targets
\[
t_{u,v,c} = \mathbb{I}(f > c), \qquad c = 1,\dots,C
\]
The reward head predicts logits $\hat{z}_{u,v,c}$ for each cutpoint.

Because logged feedback is observed under historical exposure policies, reward learning is susceptible to exposure bias. To correct for this effect, we introduce an impression propensity head that estimates the exposure probability of an item for a given user:
\[
\hat{p}_{u,v} = \sigma(\hat{z}_{\text{imp},u,v})
\]
Using this estimate, we first compute a clipped inverse propensity score and then apply self-normalization for scale stability during training.
\[
\begin{gathered}
    \tilde{w}_{u,v}
    =
    \frac{s_{u,v}}{\sum_{(u',v')\in D_r} s_{u',v'}},
    \\
    \text{where} \quad
    s_{u,v} = \min\!\left(\frac{1}{\hat{p}_{u,v}}, \tau_{\text{IPS}}\right)
    \end{gathered}
\]
This weight is applied to the ordinal loss:
\[
L_{\text{reward}}
=
\sum_{(u,v)\in D_r}
\tilde{w}_{u,v}
\sum_{c=1}^{C}
\mathrm{BCE}(\hat{z}_{u,v,c}, t_{u,v,c}).
\]
At inference time, we aggregate the ordinal outputs into a continuous reward score,
\[
R_{\phi}(u,v) = \sum_{c=1}^{C} \sigma(\hat{z}_{u,v,c})
\]
which serves as a dense utility estimate for downstream alignment.

\subsection{Multi-phase Fine-Tuning}

\subsubsection{Supervised Fine-Tuning}

The supervised adaptation stage consists of two sequential phases.

In the LP phase, the FM backbone remains frozen, and only the downstream module and prediction heads are optimized. The overall supervised objective is
\[
L_{\text{SFT}} = L_{\text{policy}} + L_{\text{reward}} + L_{\text{propensity}}
\]
This phase allows the randomly initialized downstream components to align with the pretrained FM representation space under a stable optimization landscape, thereby reducing the risk that high-variance gradients corrupt the backbone.

In the subsequent FFT phase, the FM backbone is unfrozen to enable joint specialization of all model components. Because the downstream components have already been stabilized during LP, the optimization is better conditioned and less prone to catastrophic forgetting. To further stabilize joint training, we use a discriminative learning-rate strategy: the FM backbone is updated with a smaller learning rate than the downstream modules, typically by one order of magnitude. This allows the task-specific components to adapt rapidly while the pretrained backbone co-adapts more conservatively.

\begin{algorithm}[t]
\small
\SetAlgoLined
\DontPrintSemicolon

\caption{Three-Phase Progressive Post-Training}
\label{alg:multi_phase_rft}

\SetKwInput{Input}{Input}
\SetKwInput{Output}{Output}

\Input{Pretrained foundation model $E_{\mathrm{FM}}$, downstream dataset $\mathcal{D}$}
\Output{Serving policy $\pi_\theta$, reward model $R_\phi$}

\BlankLine
Initialize downstream module $E_{\mathrm{DS}}$, serving policy head $\pi_\theta$, reward head $R_\phi$, and propensity head $P_\psi$\;

\BlankLine
\tcp{\textbf{Phase 1: Linear Probing(LP)}}
Freeze $E_{\mathrm{FM}}$\;
\ForEach{batch $(u,v,y_{\mathrm{click}},y_{\mathrm{funnel}},y_{\mathrm{imp}})\in\mathcal{D}$}{
    Compute fused representation from $E_{\mathrm{FM}}$ and $E_{\mathrm{DS}}$\;
    $\mathcal{L}_{\mathrm{policy}} \leftarrow \mathrm{CE}(\pi_\theta(u,v), y_{\mathrm{click}})$\;
    $\mathcal{L}_{\mathrm{prop}} \leftarrow \mathrm{BCE}(P_\psi(u,v), y_{\mathrm{imp}})$\;
    Compute SNIPS weight $\tilde{w}_{u,v}$ from $P_\psi(u,v)$\;
    $\mathcal{L}_{\mathrm{reward}} \leftarrow \tilde{w}_{u,v}\cdot \mathrm{OrdinalBCE}(R_\phi(u,v), y_{\mathrm{funnel}})$\;
    Update $E_{\mathrm{DS}}$, $\pi_\theta$, $R_\phi$, $P_\psi$ by minimizing
$\mathcal{L}_{\mathrm{policy}}+\mathcal{L}_{\mathrm{reward}}+\mathcal{L}_{\mathrm{prop}}$\;
}

\BlankLine
\tcp{\textbf{Phase 2: Full Fine-Tuning(FFT)}}
Unfreeze $E_{\mathrm{FM}}$\;
\ForEach{batch $(u,v,y_{\mathrm{click}},y_{\mathrm{funnel}},y_{\mathrm{imp}})\in\mathcal{D}$}{
    Compute fused representation from $E_{\mathrm{FM}}$ and $E_{\mathrm{DS}}$\;
    Compute $\mathcal{L}_{\mathrm{policy}}, \mathcal{L}_{\mathrm{reward}}, \mathcal{L}_{\mathrm{prop}}$\;
    Update all parameters by minimizing
    $\mathcal{L}_{\mathrm{policy}}+\mathcal{L}_{\mathrm{reward}}+\mathcal{L}_{\mathrm{prop}}$,
    using a smaller learning rate for $E_{\mathrm{FM}}$\;
}

\BlankLine
\tcp{\textbf{Phase 3: GRPO-based RFT}}
Initialize reference policy $\pi_{\mathrm{ref}} \leftarrow \pi_\theta$\;
Freeze $R_\phi$ and $P_\psi$\;
\ForEach{user state $u$}{
    Greedily select top-$K$ candidate items $\mathcal{A}_K$ using $\pi_\theta(\cdot\mid u)$\;
    \ForEach{$a_k\in\mathcal{A}_K$}{
        $r_k \leftarrow R_\phi(u,a_k)$\;
    }
    Compute normalized advantages
    $A_k \leftarrow \big(r_k-\bar{r}\big)/\sigma_r$\;
    Compute policy ratio
    $\rho_k(\theta)\leftarrow \pi_\theta(a_k\mid u)/\pi_{\mathrm{ref}}(a_k\mid u)$\;
    Compute GRPO objective $L(\theta)$~(Eq.~\ref{eq:grpo_loss})\;
    Update $\pi_\theta$ and $E_{\mathrm{FM}}$ by minimizing $L(\theta)$,
    using a smaller learning rate for $E_{\mathrm{FM}}$\;
}
\end{algorithm}

\subsubsection{Reinforcement Fine-Tuning}

Having stabilized the task-specific predictors and calibrated the debiased reward model, the framework then transitions to policy optimization. A snapshot of the model at the end of FFT is preserved as the frozen reference policy $\pi_{\text{ref}}$. During this stage, the policy head and FM backbone is optimized while the reward and propensity models remain frozen.

\paragraph{GRPO Formulation} For each user state, the agent selects a group of candidate actions $\{a_k\}_{k=1}^{K}$ from the current policy and receives dense scalar feedback from the reward model $R_{\phi}(a_k)$. By incorporating exposure propensity correction, the reward estimate remains consistent under potential exposure bias, including items infrequently surfaced by historical policies. The advantage is computed by normalizing the reward within the group:
\[
\begin{gathered}
 A_k = \frac{R_{\phi}(a_k) - \bar{R}}{\sigma_R},
 \\
 \text{where} \quad \bar{R} = \frac{1}{K} \sum_{k=1}^{K} R_{\phi}(a_k), \quad \sigma_R = \text{std}\big(R_{\phi}(a_k)\big)
\end{gathered}
\]

The policy ratio with respect to the frozen reference is $\rho_k(\theta) = \pi_{\theta}(a_k) / \pi_{\text{ref}}(a_k)$, and the clipped surrogate objective is:
\[
L_k(\theta) = \min\Big(\rho_k(\theta)\, A_k,\; \text{clip}\big(\rho_k(\theta),\, 1{-}\epsilon,\, 1{+}\epsilon\big)\, A_k\Big)
\]

The overall GRPO objective with KL regularization is:
\begin{equation}
\label{eq:grpo_loss}
L(\theta) = -\mathbb{E}_{k}\left[L_k(\theta)\right] + \beta \cdot D_{\text{KL}}\big(\pi_{\theta}(\cdot \mid s) \,\|\, \pi_{\text{ref}}(\cdot \mid s)\big)
\end{equation}

\paragraph{DPO Formulation} Alternatively, the reward model constructs preference pairs for direct policy optimization without explicit human annotations. For a candidate group, the preference set is defined as $\mathcal{P} = \{(a_i, a_j) \mid R_{\phi}(a_i) > R_{\phi}(a_j)\}$. For each pair, the log-probability difference under the current and reference policies is $\Delta_k = \log \pi_{\theta}(a_k) - \log \pi_{\text{ref}}(a_k)$, and the policy is optimized by:
\[
L_{\mathrm{DPO}}(\theta) = -\mathbb{E}_{(a_i, a_j) \in \mathcal{P}}\left[\log \sigma\big(\beta\,(\Delta_i - \Delta_j)\big)\right]
\]

\section{Evaluation}
\subsection{Offline Evaluation}

\paragraph{Dataset}
We evaluate our framework on large-scale interaction logs collected from a production South Korean Webtoon platform. For offline evaluation, we use interaction data from the most recent three-week period. We construct the test set by randomly selecting 10\% of users and evaluating sessions sampled from their interactions within this window.

\paragraph{Foundation Model and Downstream Features}
Our recommendation backbone is a sequential foundation model built on an HSTU encoder with approximately 7M parameters. It serves as the pretrained sequential representation backbone for downstream adaptation and policy alignment.

The downstream task focuses on content discovery, where discovery is defined as recommending titles that the user has not consumed before. To support this task, we incorporate additional long-horizon features beyond the FM inputs, including behavioral statistics such as free-episode consumption, paid-episode consumption and continuation-related signals. These features are encoded by a lightweight encoder and fused with the FM representation for final prediction.

The policy head is trained to predict whether a click occurs after impression. Offline evaluation is conducted on test users using sessions from the evaluation window.

\paragraph{Training Details}
The overall optimization procedure follows the three-phase framework described in Section~\ref{sec:methodology}. For supervised adaptation, all 1-phase baselines are trained for 100 epochs. For 2-phase variants, we allocate 50 epochs to LP and 50 epochs to FFT so that the total supervised training budget remains comparable. The learning rate is set to $10^{-3}$ in LP and $10^{-4}$ in FFT. When learning-rate control is applied, the FM backbone is updated with a smaller learning rate of $10^{-5}$ during FFT.

All RFT variants are initialized from the 2-phase SFT checkpoint and trained for an additional 50 epochs. During RFT, we jointly update the policy head and the FM backbone, using a learning rate of $10^{-5}$ for the policy-related parameters and $10^{-6}$ for the FM backbone under learning-rate control. For both GRPO and DPO, we greedily select the top-32 candidate titles from the current policy to match the serving recommendation setting. In GRPO, we use a clipping parameter of $0.1$ and a KL regularization coefficient of $0.001$.

\begin{table*}[t]
\centering
\caption{Offline ablation study. We report \textbf{Rank NDCG} for full-catalog click ranking and \textbf{Funnel NDCG} for ranking impressed titles by multi-stage engagement depth.}
\Description{Table comparing Rank NDCG and Funnel NDCG across model configurations in three groups: supervised fine-tuning variants, a direct reward-model policy, and reinforcement fine-tuning variants. GRPO with a reward model scores highest on both metrics, and two-phase supervised fine-tuning outperforms single-phase variants.}
\label{tab:offline_ablation}
\begin{tabular}{lcc}
\toprule
\textbf{Model Configuration} & \textbf{Rank NDCG} & \textbf{Funnel NDCG} \\
\midrule

\multicolumn{3}{l}{\textbf{SFT for Stable Downstream Adaptation}} \\
1-Phase SFT (Only FFT) & 0.429 & 0.620  \\
1-Phase SFT (Only LP) & 0.426 & 0.620 \\
2-Phase SFT (LP $\rightarrow$ FFT w/o LR Control) & 0.434 & 0.622  \\
2-Phase SFT (LP $\rightarrow$ FFT w LR Control) & 0.437 & 0.625  \\
\addlinespace

\multicolumn{3}{l}{\textbf{Direct Reward-Based Serving Policy}} \\
Reward Model-Based Policy & 0.394 & \underline{0.637} \\
\addlinespace

\multicolumn{3}{l}{\textbf{RFT for Alignment}} \\
Baseline (2-Phase SFT, LP $\rightarrow$ FFT w/ LR Control) & 0.437 & 0.625  \\
GRPO w/ RM & \textbf{0.463} & \textbf{0.637}  \\
GRPO w/o RM (Observed Funnel Preferences) & 0.441 & 0.627 \\
DPO w/ RM & 0.423 & 0.630 \\
DPO w/o RM (Observed Funnel Preferences) & \underline{0.453} & \underline{0.637} \\
\bottomrule
\end{tabular}
\end{table*}

\subsubsection{Metrics}
We use two complementary ranking metrics. The first evaluates whether the model correctly ranks titles that receive clicks over the full catalog. The second evaluates whether the model ranks them consistently with the depth of the observed engagement funnel among titles that were actually impressed in a session. Together, these metrics assess both immediate click-ranking quality and alignment with deeper downstream engagement.

\paragraph{Rank NDCG} To measure how well the policy ranks titles that receive clicks after impression over the full title catalog, we calculate Rank NDCG. For a user state $u$, the model produces scores over the entire title set $\mathcal{V}$. We define a binary relevance vector $y \in \{0,1\}^{|\mathcal{V}|}$, where $y_v = 1$ if title $v$ is clicked in the evaluated session and $y_v = 0$ otherwise. The metric is computed as
\[
\text{RankNDCG} = \text{NDCG}(z_u, y),
\]
where $z_u \in \mathbb{R}^{|\mathcal{V}|}$ denotes the full-catalog prediction scores for user $u$.

\paragraph{Funnel NDCG} To quantify how well the model ranks impressed titles according to the depth of downstream engagement. For an evaluated session, let $\mathcal{S} \subset \mathcal{V}$ denote the set of titles that were actually impressed. For each impressed title $v \in \mathcal{S}$, we assign a graded relevance label $r_v \in \{0,\dots,L\}$ corresponding to the deepest engagement funnel stage reached by the user for that title. The metric is computed as
\[
\text{FunnelNDCG} = \text{NDCG}(z_u^{(\mathcal{S})}, r),
\]
where $z_u^{(\mathcal{S})}$ denotes the model scores restricted to the impressed titles and $r$ is the corresponding graded relevance vector.

\subsubsection{Results}
\paragraph{Phase 1--2: Supervised Fine-Tuning (LP $\rightarrow$ FFT)}
We first evaluate the progressive adaptation strategy that transitions from LP to full-parameter fine-tuning(FFT). In Phase 1, the pretrained foundation model remains frozen and only the downstream components are optimized. In Phase 2, all model parameters are trainable and jointly updated through FFT.

Table~\ref{tab:offline_ablation} compares one-phase and two-phase fine-tuning strategies. Training with a single phase only, whether LP or FFT, yields limited performance on both Rank NDCG and Funnel NDCG. In particular, 1-phase LP and 1-phase FFT both underperform the two-phase variants, indicating that neither frozen-backbone adaptation nor direct full-parameter optimization alone is sufficient for this downstream setting. By contrast, the progressive fine-tuning(LP $\rightarrow$ FFT) consistently improves both metrics, suggesting that warming up the downstream components under a frozen backbone provides a more stable starting point for subsequent full-model adaptation.

We further evaluate discriminative learning-rate control during FFT. Applying a smaller($\times 0.1$) learning rate to the pretrained FM backbone yields modest but consistent gains over the two-phase variant without learning-rate control on both Rank NDCG and Funnel NDCG. This suggests that controlling the update magnitude of the pretrained backbone helps preserve useful pretrained representations while still enabling effective downstream specialization.

\paragraph{Direct Reward-Based Serving Policy}
We next evaluate whether the learned reward model can be used directly as the serving policy. This is a particularly relevant ablation because the reward model is the component that most directly targets the business objective: it is trained to predict the engagement-funnel depth of each user--item pair. In our setting, the funnel is defined by six ordered stages: impression, click, initial episode viewing, completion of free episodes, entry into paid episodes and completion of paid content.

If business-target supervision alone were sufficient, the reward model would be a natural candidate for serving. However, Table~\ref{tab:offline_ablation} shows that directly ranking with the reward model achieves competitive Funnel NDCG but substantially worse Rank NDCG than the click-trained serving policy. This indicates that the reward model captures business-oriented engagement depth reasonably well but does not provide sufficient discrimination for full-catalog click ranking.

\paragraph{Phase 3: Reinforcement Fine-Tuning}
Starting from the 2-phase SFT checkpoint, we perform RFT to align the policy with deeper business objectives. During this stage, the policy head and the FM backbone are jointly updated, while the reward and propensity models remain fixed. To preserve the stability established during supervised adaptation, we continue to use discriminative learning-rate control, updating the FM backbone more conservatively than the task-specific components.

We compare GRPO and DPO under both reward-based and reward-free variants. In the reward-based setting, a learned reward model provides scalar utility estimates for alignment. In the reward-free setting, policy optimization is driven directly by observed funnel-based supervision. As shown in Table~\ref{tab:offline_ablation}, RFT generally improves both Rank NDCG and Funnel NDCG over the supervised baseline, indicating that post-training alignment can further refine ranking quality once the base policy has been stably adapted. Notably, GRPO and DPO with reward-based alignment raise Funnel NDCG to the level of the direct reward-model policy, while retaining substantially stronger Rank NDCG.

Among the evaluated methods, GRPO with reward modeling achieves the best overall performance, yielding the strongest gains on both metrics. This supports our central design choice: the business-target reward is most effective when used as an alignment signal for post-training rather than as a direct serving score. In particular, GRPO with reward modeling matches the Funnel NDCG of the direct reward-model policy while substantially improving Rank NDCG, showing that reward-based alignment can transfer business-oriented utility signals into the serving policy without sacrificing ranking discrimination. GRPO without reward modeling still improves over the supervised baseline, but the gains are smaller, suggesting that a learned reward provides a more informative alignment signal than observed funnel supervision alone.

DPO also improves over the supervised baseline, particularly when preference pairs are constructed directly from observed funnel signals. This indicates that explicit reward modeling is not strictly necessary for alignment when sufficiently informative graded engagement labels are available. However, in our setting, DPO with reward-model-induced preferences performs noticeably worse than both GRPO with reward modeling and DPO with observed funnel preferences. A possible explanation is that the learned reward model is better utilized as a continuous alignment signal in GRPO than as a source of synthetic pairwise preferences in DPO. Converting scalar reward estimates into binary preference pairs may discard useful relative information and amplify reward-model noise.

\begin{figure*}[t]
    \centering
    \includegraphics[width=1.0\textwidth]{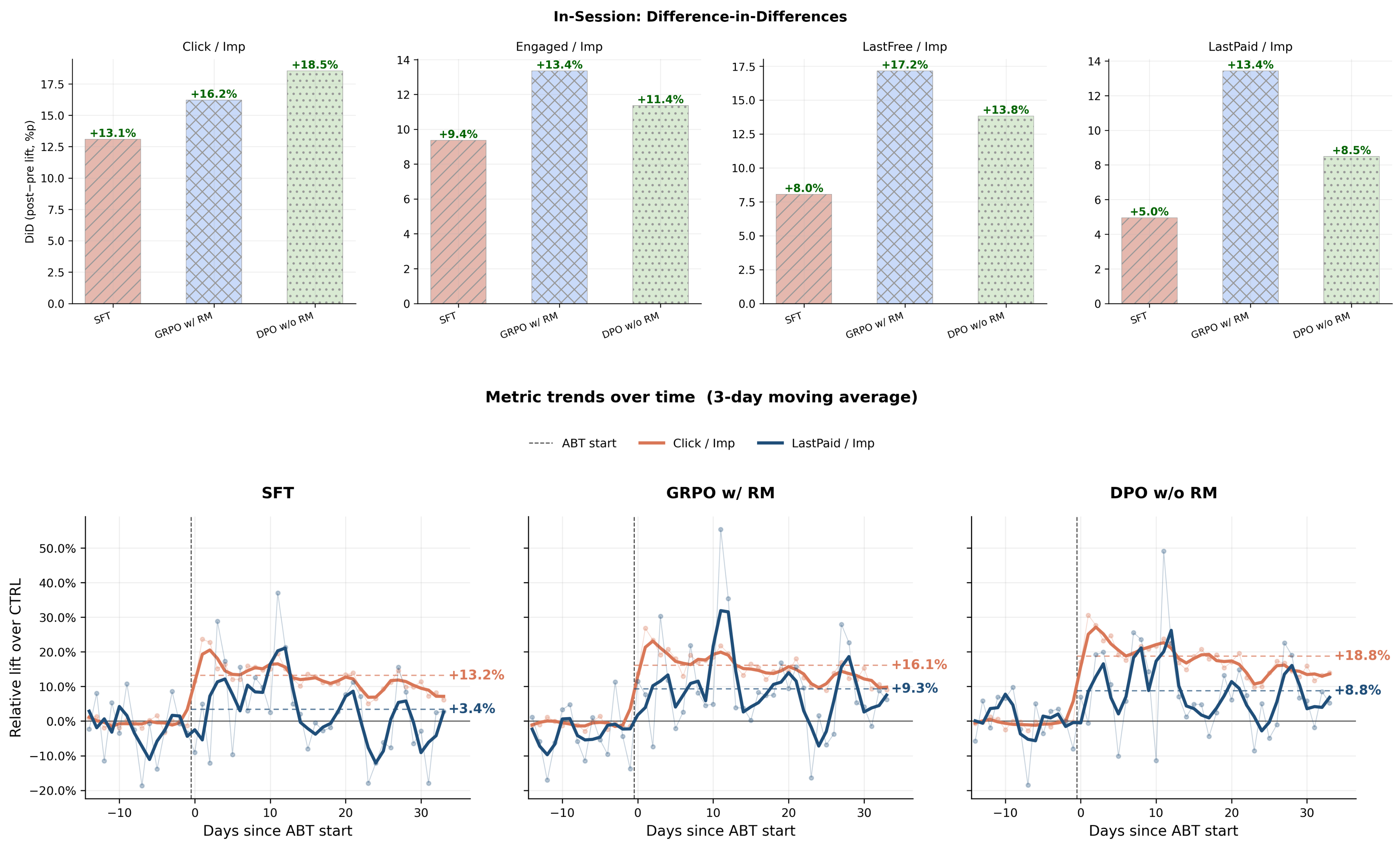}
    \caption{Relative online lift over the production control model across in-session metrics.
    The top plot shows difference-in-differences estimates per metric, and the bottom plot
    tracks the daily trajectory of Click/Imp and LastPaid/Imp per variant over the experiment
    window. All foundation-model-based variants outperform CTRL, and the aligned variants
    deliver additional gains beyond the supervised baseline.}
    \Description{Two-panel chart of online A/B test results. The top panel shows difference-in-differences lift over the control model for each in-session metric, comparing the supervised baseline, GRPO with reward model, and DPO without reward model; all bars are positive. The bottom panel plots daily lift trajectories for clicks per impression and last-paid-episode reads per impression: lift spikes right after deployment, declines over the first days, then stabilizes at a positive level, with GRPO with reward model highest on the paid metric throughout.}
    \label{fig:online_ab}
\end{figure*}

\subsection{Online Experiment}

To validate whether the offline improvements translate into practical gains in a production environment, we conducted an online A/B test on our large-scale Korean Webtoon platform.

\subsubsection{Setup}
We compared the following ranking models:
\begin{itemize}
    \item \textbf{CTRL}: the production control model without a foundation-model backbone. It follows a conventional multi-task ranking setup that separately estimates click-through rate(CTR) and conversion-related signals(CVR) and ranks candidate items by their product($\mathrm{CTCVR}$), following the standard industrial formulation of post-click conversion modeling~\cite{wang2022escm2}.

    \item \textbf{SFT}: the 2-phase supervised foundation-model baseline trained with LP $\rightarrow$ FFT and learning-rate control.

    \item \textbf{GRPO w/ RM}: the 3-phase progressive model further aligned using GRPO with the learned reward model.

    \item \textbf{DPO w/o RM}: the 3-phase progressive model aligned using DPO with observed funnel-based preferences.
\end{itemize}

\subsubsection{Metrics}
We focus on \emph{in-session per-impression} metrics, since they are directly tied to the serving objective. Specifically, we measure click, engaged reading, reaching the last free episode and reaching the last paid episode within the same session after impression. These metrics reflect not only immediate interaction but also how far recommendation exposure leads users along the downstream consumption funnel during serving.

\subsubsection{Results}
Figure~\ref{fig:online_ab} summarizes the relative lift of each model over the production control model on in-session metrics. Overall, all foundation-model-based variants outperform the control group consistently, indicating that the pretrained sequential backbone, together with progressive downstream adaptation, provides a stronger basis for recommendation than the existing CTR/CVR product ranking framework.

The 2-phase SFT baseline already yields clear improvements over CTRL on every reported metric. This suggests that replacing the non-foundation production model with a progressively adapted foundation model is beneficial even before applying reward-based alignment.

The aligned variants provide additional gains beyond SFT. GRPO with reward modeling shows stronger improvements on deeper engagement metrics, particularly \textit{Engaged/Imp} and \textit{LastFree/Imp}, while also improving click and paid-consumption outcomes. This pattern is consistent with the intended role of reward-based alignment, which is to steer the serving policy toward titles associated with deeper downstream consumption rather than optimizing only immediate clicks.

DPO without reward modeling also shows better performance than SFT. In our experiment, it achieves the largest gain on \textit{Click/Imp}, while remaining competitive on deeper funnel metrics. Compared with GRPO w/ RM, its improvements are relatively more pronounced on immediate interaction, whereas GRPO w/ RM is slightly stronger on mid-funnel progression.

Figure~\ref{fig:online_ab} also shows the temporal evolution of online lift in \textit{Click/Imp} and \textit{LastPaid/Imp} over the one-month A/B test period. All foundation-model-based variants exhibit elevated lift immediately after deployment, followed by a gradual decline during the first several days. This pattern is commonly observed in online recommendation experiments. After this transient period, the lift stabilizes at a lower but significantly positive level across all variants and metrics (p < 0.001). A key observation is that the relative ordering among variants remains stable throughout the experiment. On \textit{LastPaid/Imp}, GRPO w/ RM consistently shows the highest lift, followed by DPO w/o RM and then SFT. This indicates that reward-based alignment yields more persistent gains on deeper engagement outcomes.

\subsubsection{Discussion}
The online results support two main observations. First, the foundation-model-based SFT baseline consistently improves over the non-foundation production control on all reported in-session metrics. Second, reinforcement-based post-training provides additional gains beyond SFT, suggesting that alignment signals derived from downstream business objectives can be effectively transferred to the serving policy in a production environment.

The difference between GRPO w/ RM and DPO w/o RM further suggests that the choice of alignment method influences which part of the in-session funnel is emphasized. In our experiment, GRPO w/ RM yields stronger gains on mid-to-deep engagement metrics (\textit{Engaged/Imp} and \textit{LastFree/Imp}), whereas DPO w/o RM achieves the largest gain on immediate click behavior (\textit{Click/Imp}).

This pattern may reflect differences both in the form of supervision and in how that supervision is incorporated into policy updates. GRPO directly optimizes against the continuous scalar outputs of the learned reward model. Through advantage-based updates, it can use not only the ordering of candidate items but also the relative magnitude of their estimated utility, which may be beneficial for improving downstream engagement depth.

By contrast, DPO operates on pairwise preferences. When continuous reward scores are converted into binary preference pairs, information about the magnitude of the reward difference is no longer preserved. In addition, small and potentially noisy score differences must be converted into hard comparisons, which can make the resulting preference signal less reliable. This is consistent with the relative advantage of GRPO w/ RM observed in our results.

When DPO constructs preference pairs directly from observed funnel outcomes, however, it avoids reliance on the learned reward model. Although such supervision is sparser, it provides direct and relatively unambiguous preference signals from realized user behavior. This may make DPO particularly effective for sharpening top-of-funnel ranking decisions, such as click-oriented selection.

\section{Conclusion}

In this work, we proposed a three-phase progressive post-training framework for adapting and aligning recommender foundation models with practical business objectives. By explicitly separating task-specific knowledge adaptation from reward-driven alignment, the framework provides a structured way to reduce the instability of direct full-parameter optimization and to bridge the gap between dense implicit-feedback supervision and sparse business-oriented targets.

Both offline and online evaluations support the effectiveness of this progressive design. The transition from LP to FFT provides a stable adaptation path, enabling downstream specialization while preserving useful pretrained representations. Building on this stabilized policy, RFT further improves alignment with business-oriented objectives. In particular, our results suggest that using a reward model as an alignment signal is more effective than directly deploying the reward model itself as the serving policy.

Our large-scale online experiments further show that the proposed framework improves production recommendation quality over a conventional non-foundation baseline across in-session discovery metrics. The comparison between alignment methods also suggests that different post-training objectives may emphasize different parts of the engagement funnel: GRPO with a learned reward model is more effective for deeper engagement progression, whereas DPO with observed funnel preferences remains highly competitive on immediate interaction metrics.

Overall, these findings indicate that progressive post-training is a practical and effective approach for deploying foundation models in industrial recommender systems, and that explicitly separating supervised adaptation from reward-based alignment can improve both optimization stability and downstream serving performance.

\bibliographystyle{ACM-Reference-Format}
\bibliography{refs}

\end{document}